\documentclass[12pt]{article}

\usepackage{amsthm,amsmath}
\usepackage{natbib}
\setcitestyle{authoryear,open={((},close={))}}
\usepackage{booktabs}
\usepackage{articlestyle}
\usepackage{authblk}
\usepackage{marginnote} %Package helping with margin notes

\usepackage{mathtools}

\graphicspath{{../figs/},{figs/}}

\usepackage{url}
\usepackage{ulem}
\usepackage[group-separator={,}]{siunitx}
\usepackage{xr}
\usepackage{color}

\title{Latent modeling of flow cytometry cell populations}
\author[1]{Jonas Wallin}
\author[2]{Kerstin Johnsson\thanks{johnsson@maths.lth.se; Corresponding author}}
\author[2,3]{Magnus Fontes}
\affil[1]{Mathematical Sciences, Chalmers and University of Gothenburg}
\affil[2]{Centre for Mathematical Sciences, Lund University}
\affil[3]{Institut Pasteur, Paris}

\begin{document}

\maketitle

\begin{abstract}
Flow cytometry is a widespread single-cell measurement technology with a multitude of clinical and research applications. Interpretation of flow cytometry data is hard; the instrumentation is delicate and can not render absolute measurements, hence samples can only be interpreted in relation to each other while at the same time comparisons are confounded by inter-sample variation. Despite this, current automated flow cytometry data analysis methods either treat samples individually or ignore the variation by for example pooling the data. In this article we introduce a Bayesian hierarchical model for studying latent relations between cell populations in flow cytometry samples, thereby systematizing inter-sample variation. The model is applied to a data set containing replicated flow cytometry measurements of samples from healthy individuals, with informative priors capturing expert knowledge. It is shown that the technical variation in the inferred cell population sizes is small in comparison to the intrinsic biological variation. The large size of flow cytometry data, where a single sample can contain measurements on hundreds of thousands of cells, necessitates computationally efficient methods. To address this, we have implemented a parallel Markov Chain Monte Carlo scheme for sampling the posterior distribution. 
\end{abstract}

\section{Introduction}
In a flow cytometer a number of characteristics for each individual cell in a sample of $\sim \!\! 10^4$ to $\sim \!\! 10^6$ cells are quantified as they pass through it in a fluid stream. The data that are obtained are most often summarized by grouping cells into cell populations; properties of these cell populations are used in many clinical applications---for example monitoring HIV infection and diagnosing blood cancers---and in many branches of medical research (\citealp{shapiro05,nolan07}). Defining the cell populations based on the measured characteristics is in state-of-the-art analyses still done manually by trained operators looking at two-dimensional projections of the data. However, the importance of automated methods has risen along with an increase of the dimension of typical flow cytometry data sets due developments in flow cytometry technology (\citealp{oneill13}) and the emergence of studies with large numbers of flow cytometry samples (\citealp{chen15}).

Automatic cell population identification is hard since flow cytometry measurements are not absolute, while at the same time different samples cannot be directly compared due to technical variation---especially apparent when samples are analyzed at different laboratories (\citealp{welters12})---and intrinsic biological variation. Despite this, research into automated population identification methods has focused on individual or pooled flow cytometry samples, sometimes attempting to align data at first through normalization procedures (\citealp{hahne10}). We introduce a Bayesian hierarchical model with latent relations between flow cytometry samples, thereby allowing for a systematic study of variation of cell population characteristics in a collection of samples and additionally enabling the use of prior information about cell populations. 

Splitting the cell measurements in a sample into cell populations is essentially a clustering problem. In the context of flow cytometry data analysis clustering is called automatic gating, as opposed to the manual gating performed by operators. Model-based clustering using mixture models has been the most used approach for automated gating (\citealp{lo08,boedigheimer08,chan08,pyne2009automated,hu13,naim14}). Mixture models are very well suited to describe flow cytometry data because they have a natural biological interpretation based on the cell populations. Examples of other approaches that have been used for automated gating are grid based density clustering (\citealp{qian10}), spectral clustering (\citealp{zare10}), hierarchical clustering (\citealp{qiu11,bruggner14}) and k-means clustering (\citealp{aghaeepour11,ge12}). An evaluation of a wide range of automated gating methods was performed in the FlowCAP I challenge (\citealp{aghaeepour13}). No method clearly outperformed the others.

Apart from pooling data, two approaches for identifying cell populations jointly in a collection of flow cytometry samples have been put forward previously. The first one is to cluster each sample separately and then match the resulting clusters (\citealp{pyne2009automated,azad13}). The second one is a Dirichlet process Bayesian hierarchical model ignoring variation in location and shape between cell populations (\citealp{cron13}). This limitation precludes analysis of any such latent variation.

In our model, the cells in a sample are clustered using a multivariate Gaussian mixture model (GMM), where $K$ components describe cell populations and one component describes outliers. Outliers can arise for example from dead cells, non-specific binding of markers and doublets, i.e.\ pairs or groups of cells that pass through the flow cytometer at the same time. For each component not representing outliers its mean and covariance matrix is linked to a latent cluster which collects corresponding components across all samples. In practice this is done by assuming a normal prior for the means and an inverse Wishart prior for the covariance matrices of the components linked to a given latent cluster. The variation in location and shape between corresponding mixture components across samples is controlled by the priors on parameters of the latent clusters. The location of component means and shape of components can also be restricted if there is prior information supporting this. The probabilistic assumptions of the model are formulated in Section \ref{sec:model}. In an extension of the model we include the possibility to account for that not all populations are present in every sample, which is a frequent situation in flow cytometry data sets which is challenging for joint analyzes.  

Our model differs from previous Bayesian hierarchical models of mixtures (\citealp{Lopes2003,Muller2004,teh06,salakhutdinov10}) in that the mixture components forming the Gaussian mixture model for one sample are non-exchangeable; they represent different cell populations and thus have different priors. Furthermore, latent relations between the mean and covariance parameters of mixture components have previously only been studied in the case when the covariance matrices were assumed to be diagonal (\citealp{salakhutdinov10}).

Another challenge that has to be addressed when analyzing flow cytometry data is that cell populations in flow cytometry data can be skew and/or have heavy tails and are then not fit well by a Gaussian component (\citealp{lo08, pyne2009automated, Fruhwirth2010skewt}). To handle this we use multiple components to describe such populations, an approach that have often been employed for flow cytometry data (\citealp{finak09,chan08,baudry10,naim14}) and has the further advantage that the number of cell populations can be automatically detected. We merge Gaussian components into super components with a procedure based on a systematic study of methods for merging mixture components (\citealp{hennig10}); details are given in Section \ref{sec:merging}.

In Section \ref{sec:SimDat} we fit the model to a simulated data set to evaluate the ability of the sampling scheme to recover the model parameters. In Section \ref{sec:realworld} we apply our model to a real world flow cytometry data set where we detect both known cell populations with prior information on population locations and cell populations without this prior information. We then apply Principal Component Analysis (PCA) to the inferred cell population sizes and find that biological variation between individuals can be separated from technical variation between replicates.

\section{Methods}

\subsection{Model}
\label{sec:model}
% % % % % % % % % % % % % % % % % % % % % % % % % % % % % % % % % %
%
%	hierarchical Gaussian
%
%
%
%
% % % % % % % % % % % % % % % % % % % % % % % % % % % % % % % % % % %
%In this section we formulate the probabilistic assumptions of the model. 
Let $\mv{Y}_{ij}$ denote vector valued measurement number $i$ in sample $j$. Here  $i\in \{1,\dotsc,n_j\}$,  where $n_j$ is the number of cells in sample $j$, and $j\in\{1,\dotsc,J\}$, where $J$ is the number of samples. We let the dimension of the observations be denoted $d$. With $K$ components describing cell populations the probability density for cell measurement $i$ of a flow cytometry sample $j$ is modeled as 
\begin{equation} 
f(\mv{Y}_{ij}) = \sum_{k = 1}^K \pi_{jk}N(\mv{Y}_{ij}; \mv{\mu}_{jk}, \mv{\Sigma}_{jk}) + \pi_{j0}N(\mv{Y}_{ij};\mv{\mu}_{j0},\mv{\Sigma}_{j0}), \label{eq:mixmod} 
\end{equation}
where $N(\mv{Y}; \mv{\mu}, \mv{\Sigma})$ denotes the probability density function of the normal distribution with mean $\mv{\mu}$ and covariance matrix $\mv{\Sigma}$ evaluated at $\mv{Y}$. The number $K$ is usually chosen to be much larger than the expected number of cell populations since for non-Gaussian cell populations many components are needed to describe them. The last component represents outliers and its parameters $\mv{\mu}_{j0} = \mv{\mu}_0$ and $\mv{\Sigma}_{j0} = \mv{\Sigma}_0$ are identical across samples. The vector $\mv{\pi}_j = \{\pi_{j0},\dotsc,\pi_{jK}\}$ contains the mixing proportions, i.e.\ the proportion of cells described by the component. To connect the cell populations between samples we use a latent layer, assuming that for a given $k$ each $\mv{\mu}_{jk}$ and $\mv{\Sigma}_{jk}$ is drawn from a normal and an inverse Wishart distribution respectively. Specifically, in our model, for $k = 1,\dotsc,K$,
\begin{equation} \label{eq:latent}
\mv{\mu}_{jk} | \mv{\theta}_{k}, \mv{\Sigma}_{\theta_k} \sim N(\mv{\theta}_{k}, \mv{\Sigma}_{\theta_k}), \quad
\mv{\Sigma}_{jk}| \mv{\Psi}_k,\nu_k \sim IW(\mv{\Psi}_k,\nu_k)
\end{equation} 
where $\mv{\theta}_k$, $\mv{\Sigma}_{\mu_k}$, $\mv{\Psi}_k$ and $\nu_k$ are hyper-parameters describing latent cluster $k$. The main reason for using the normal and inverse Wishart distributions is computational efficiency, since these are conjugate priors to the mean and the covariance respectively of the normal distribution. We call $\mv{\theta}_k$ and $\mv{\Psi}_k/(\nu_k-d-1)$ the latent cluster mean and latent cluster covariance matrix respectively, since they are the a priori expected values of $\mv{\mu}_{jk}$ and $\mv{\Sigma}_{jk}$.

\begin{figure}[tb]
	\centering
	\includegraphics[width=.6\textwidth]{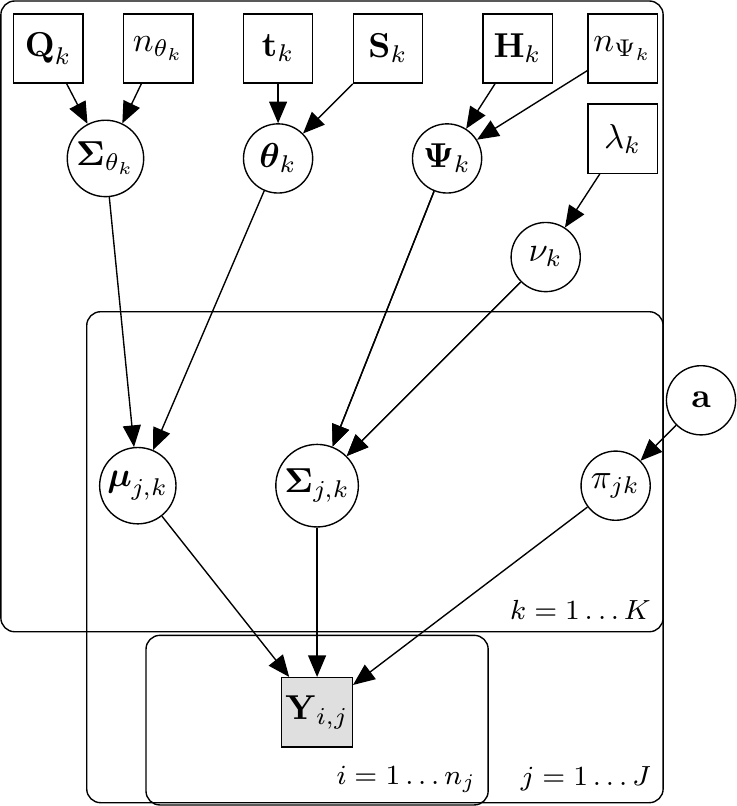}
	\caption{Directed acyclic graph describing the Bayesian hierarchical model. Square boxes indicate that the values are known.}
	\label{fig:dag}
\end{figure}

For the hyper-parameters describing the latent clusters and the mixing proportions we use the following prior distributions:
\begin{align} \label{eq:priors}
 \mv{\theta}_k | \mv{t}_k,\mv{S}_k &\sim  N(\mv{t}_k,\mv{S}_k),  &  \mv{\pi}_j &\sim \, D(\mv{a}),\\
\mv{\Sigma}_{\theta_k}|\mv{Q}_k,n_\theta &\sim IW(\mv{Q}_k,n_\theta),  &  \nu_k|\lambda_k &\sim  \exp(-\lambda_k), \nonumber \\
\mv{\Psi}_k |\mv{H}_k,n_\Psi  &\sim W(\mv{H}_k,n_\Psi),  & &\nonumber
\end{align}
where $W$ denotes the Wishart distribution and $D$ denotes the Dirichlet distribution, which is conjugate prior to the multinomial distribution. For each $\nu_k$ we assign a exponential prior on the positive natural numbers. The complete structure of the model is displayed through a DAG in Fig.\ \ref{fig:dag}.

The parameters $\mv{t}_k$ and $\mv{S}_k$ define the prior belief of the location of the latent means $\mv{\theta}_k$, whereas the parameters $\mv{Q}_k$ and $n_\theta$ control the spread of component means within a latent cluster and are hence important to control the variation between samples. A large $n_{\theta_k}$ along with a small $\mv{Q}_k$ forces the $\mv{\mu}_{jk}$ together; it makes large deviations between $\mv{\Sigma}_{\theta_k}$ and $\mv{Q}_k$ unlikely. On the other hand, the parameters $\mv{H}_k$ and $n_{\Psi_k}$ control the latent covariance matrices and the variation between component covariance matrices. If $n_{\Psi_k}$ is large each $\mv{\Sigma}_{jk}$ will be close to $\mv{\Psi}_k / (\nu_k-d-1)$ for any $k$; note that a high $n_{\Psi_k}$ makes high $\nu_k$ more probable.

Finally, to simplify sampling from the posterior distribution of the parameters, we add an component assignment variable $x_{ij} \in \{0,1,\dotsc,K\}$ describing which component $\mv{Y}_{ij}$ is drawn from. To comply with (\ref{eq:mixmod}), the a priori uncertainty of component membership is modeled by $ x_{ij} \sim \, \Mult(\mv{\pi}_{j},1)$,
where $\Mult$ denotes the multinomial distribution.

The resulting posterior distribution of all the parameters, denoted $\mv{\Theta}$, and $\mv{x}$ given the data $\mv{Y}$ is given in the Supplemental material, Appendix \ref{sec:SMposterior}. In Appendix \ref{sec:sampling} we describe the Markov chain Monte Carlo (MCMC) sampling scheme used to generate posteriors for our model parameters.

\subsubsection{Absent components}
In some flow cytometry data sets not all cell populations are present in all samples. In our model this corresponds to that $\pi_{jk} = 0$ for some $(j,k)$. However, mixture component parameters for empty clusters will still affect the mixing of the MCMC for the parameters of the latent cluster. It can also happen that if a cluster is empty that the mixture component moves and split a neighboring cluster in two. To avoid this in such data sets we extend our model by introducing a variable $\mv{Z}_j \in \{0,1\}^K$ that says which components are active in sample $j$. This has the further advantage that when sampling from the posterior distribution of the model we get the probability for each cluster that it is present in a sample. We impose a prior on $\mv{Z}_j$ which is proportional to $\exp(-c_s\sum_{k=1}^K\mv{Z}_j) I(\sum_{k=1}^K\mv{Z}_j>0)$  where $I$ denotes the indicator function and $c_s>0$. The prior makes the model prefer fewer activated clusters so that if there is a very small cluster the likelihood will be larger if it is inactivated.

The changes to (\ref{eq:mixmod})--(\ref{eq:priors}) required by this extension are straightforward but inference of the model becomes a bit more involved since removing components reduces the dimension of the model. To accommodate for this we have included a reversible jump step in our sampling algorithm. Details are given in the Supplemental Material, Appendix \ref{sec:sampling}.

\subsection{Merging latent clusters} \label{sec:merging}
To determine the ``correct'' number of clusters in a data set directly from the data is an ill-defined problem, since what should be considered to be a separate cluster depends on the interpretation of the data. Nevertheless, there are many different criteria which can be used to guide the decision about the number of populations (\citealp{fruhwirth06,hennig10}). We use overlap between components---measured by Bhattacharyya distance---and unimodality of the resulting super clusters---measured by Hartigan's dip test (\citealp{hartigan85})---to determine which latent clusters to merge and to indicate our confidence in the mergers. 

In an evaluation of criteria for merging Gaussian components, the Bhattacharyya distance performed well (\citealp{hennig10}). Bhattacharyya distance merges clusters according to a pattern-based cluster concept as opposed to a modality-based concept (\citealp{hennig10}), meaning that a small dense cluster inside a large sparse cluster will not be merged into the sparse cluster when their densities are disparate, even when the resulting super cluster would have had a unimodal density. This makes sense for flow cytometry data since two such clusters could very well describe two different cell populations.

The Bhattacharyya distance between $N(\mv{\mu}_1,\mv{\Sigma}_1)$ and $N(\mv{\mu}_2,\mv{\Sigma}_2)$ is 
\begin{equation} \label{eq:bhat}
 d_{\text{bhat}} = 1/8 \cdot (\mv{\mu}_1 - \mv{\mu}_2)^\trsp \bar{\mv{\Sigma}}^{-1} (\mv{\mu}_1 - \mv{\mu}_2) + 1/2 \cdot \log \left(|\bar{\mv{\Sigma}}| / \sqrt{|\mv{\Sigma}_2||\mv{\Sigma}_2|} \right), 
\end{equation}
where $\bar{\mv{\Sigma}} = (\mv{\Sigma}_1 + \mv{\Sigma}_2)/2$ (\citealp{fukunaga90}). In order to measure Bhattacharyya distance between mixtures of Gaussian distributions, which is necessary for deciding if super clusters should be merged with other clusters, we approximate each mixture with a Gaussian distribution. The means and the covariance matrices are estimated using a soft clustering of the data points inferred from the sampling of $x_{ij}$, detailed in the Supplementary material, Appendix \ref{sec:SMmerging}.

However, it is not obvious how to set a threshold for $d_{\text{bhat}}$, since the appropriate threshold depends on the distribution of the data (\citealp{hennig10}), which is unknown. Because of this we use a low soft threshold $d_1$ and a high hard threshold $d_2$. Two clusters closer to each other than $d_1$ are always merged, two clusters whose distance is between $d_1$ and $d_2$ are only merged if they fulfill an additional criterion based on Hartigan's dip test for unimodality.

Unimodality is an appealing heuristic for defining cell populations, and it has frequently been used for automated gating (\citealp{chan08,ge12,naim14}). It has two main limitations. The first one, populations which should be separate based on that they have disparate densities, even though they are colocalized, can be bypassed by combining unimodality with a pattern-based merging criterion such as Bhattacharyya distance. The second one, that it is difficult to determine if a multi-dimensional empirical distribution is multimodal, is usually handled by considering one-dimensional projections (\citealp{hennig10,naim14}). This is the approach we take here, using Hartigan's dip test of unimodality for each of the projections onto the coordinate axis and for the projection onto Fisher's discriminant coordinate. If for a proposed merger, any of these projections is found to be multimodal, the clusters are not merged. Further details of the merging procedure is given in the Supplemental material, Appendix \ref{sec:SMmerging}.

\section{Experiments}
\subsection{Simulated data}
\label{sec:SimDat}
In other to test the performance of the proposed sampling scheme, we use it on a simulated data set. The dimension of the simulated data is, for visual reasons, set to three. We use four latent clusters and generate eighty artificial flow cytometry samples. Each sample has measurements of \num{15000} cells. In order to test the ability to detect if a population is present in a sample or not, one of the latent clusters is present in only eight samples and one latent cluster is present in twentyfour samples. The cluster which is present in only eight samples represents a small cell population, containing  $1\%$ of the total number of cells. The parameters and the algorithm used for generating the data are given in the Supplementary material, Appendix \ref{sec:simgen}. 

\begin{figure}[tb]
	\begin{center}
		\begin{minipage}[b]{0.45\textwidth}
			\centering
			\includegraphics[width=\textwidth]{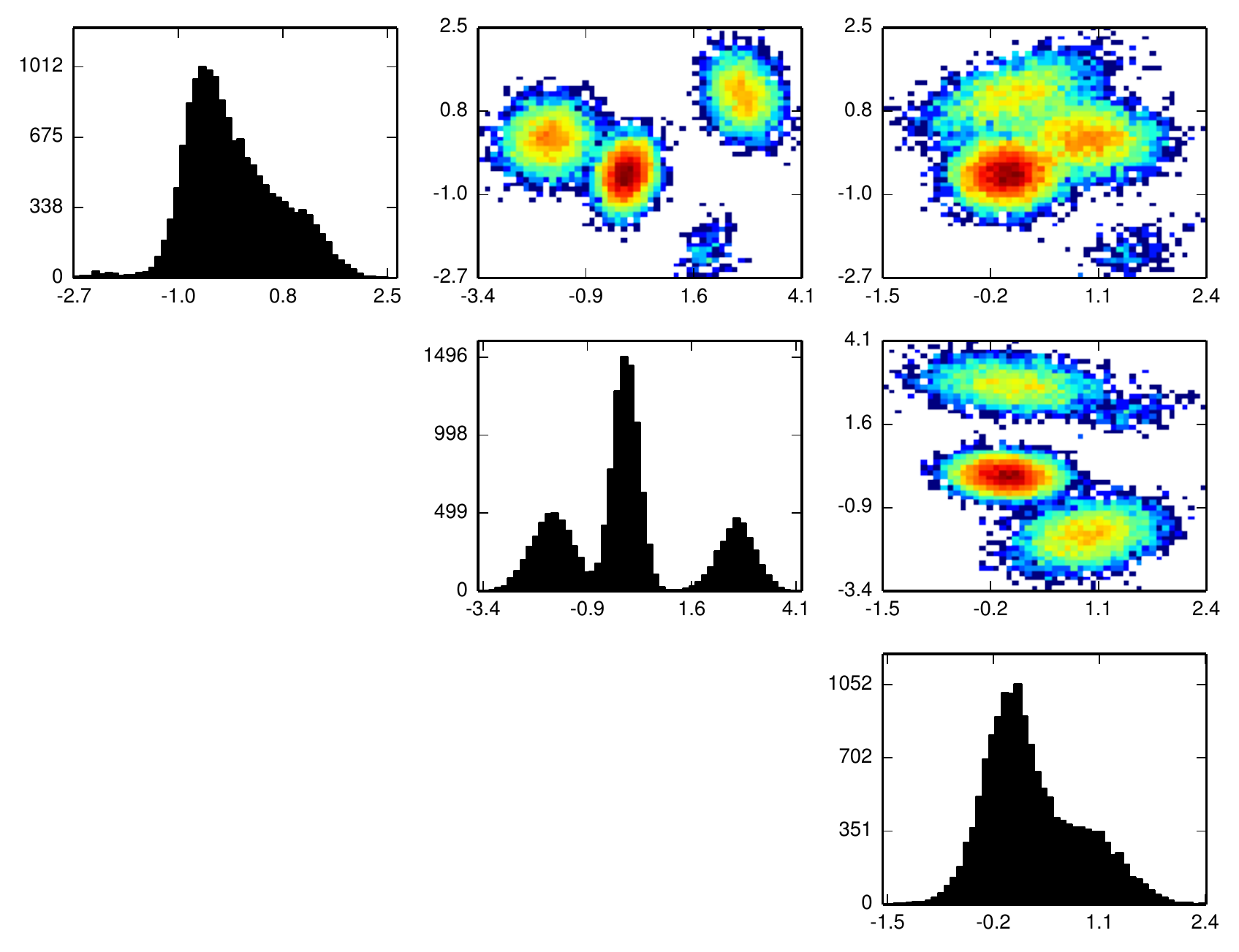}\\
			(a)
		\end{minipage}
		\begin{minipage}[b]{0.45\textwidth}
			\centering
			\includegraphics[width=\textwidth]{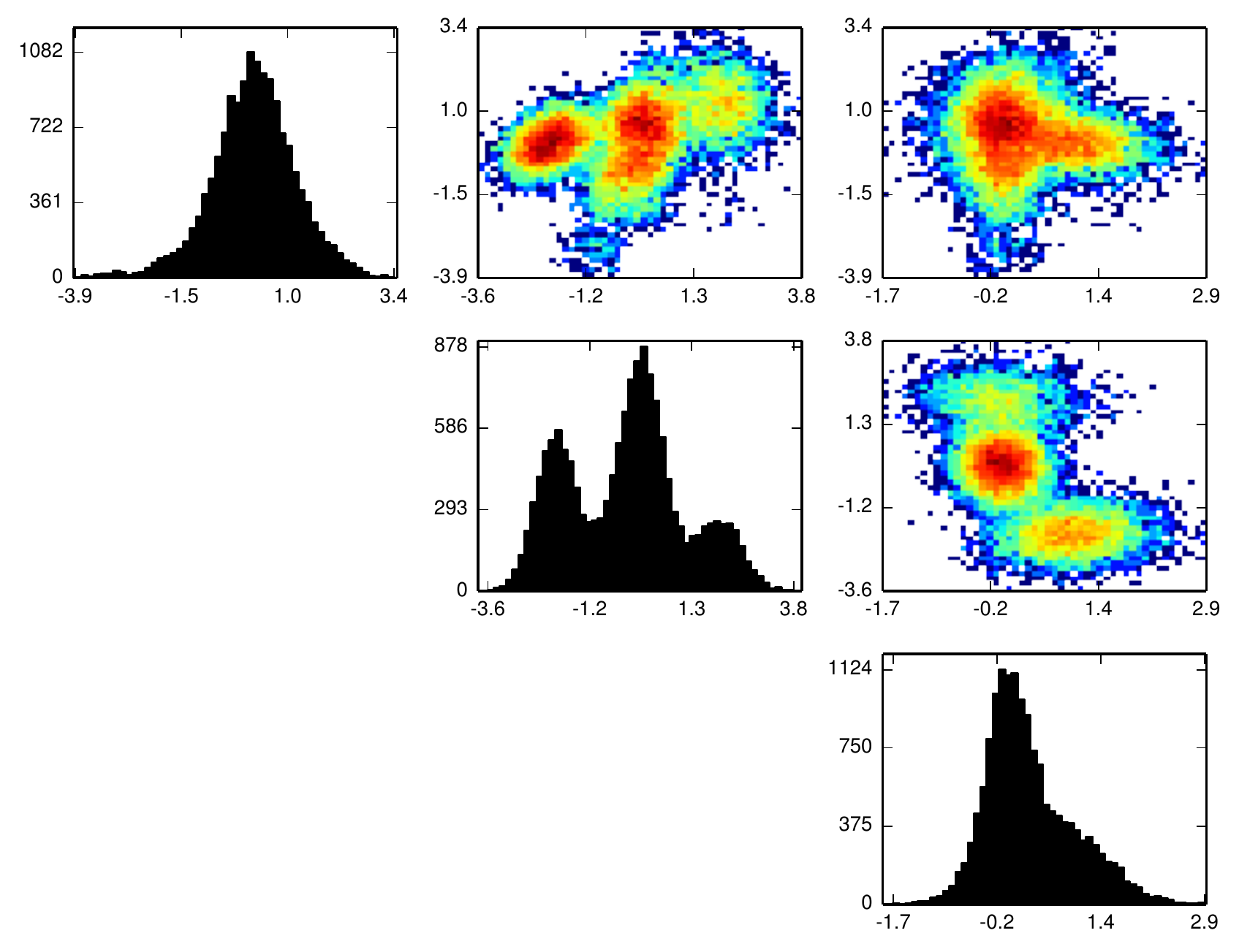} \\
			(b)
		\end{minipage}
	\end{center}
	\caption{One and two dimensional histograms for one flow cytometry sample containing \num{15000} in (a), and histograms of  \num{15000} data points drawn uniformly from the pooled data in (b).}
	\label{fig:hist}
\end{figure}

In Fig.\ \ref{fig:hist} we show univariate and bivariate histograms of all cell measurements pooled together, as well as the corresponding histograms of the data from a single sample where all four clusters are present. Note that the data when pooled together has a complicated density, as it is in fact a mixture of 232 multivariate normal densities.

\begin{figure}[tb]
	\begin{center}
		\begin{minipage}[b]{0.45\textwidth}
			\centering
			\includegraphics[width=\textwidth]{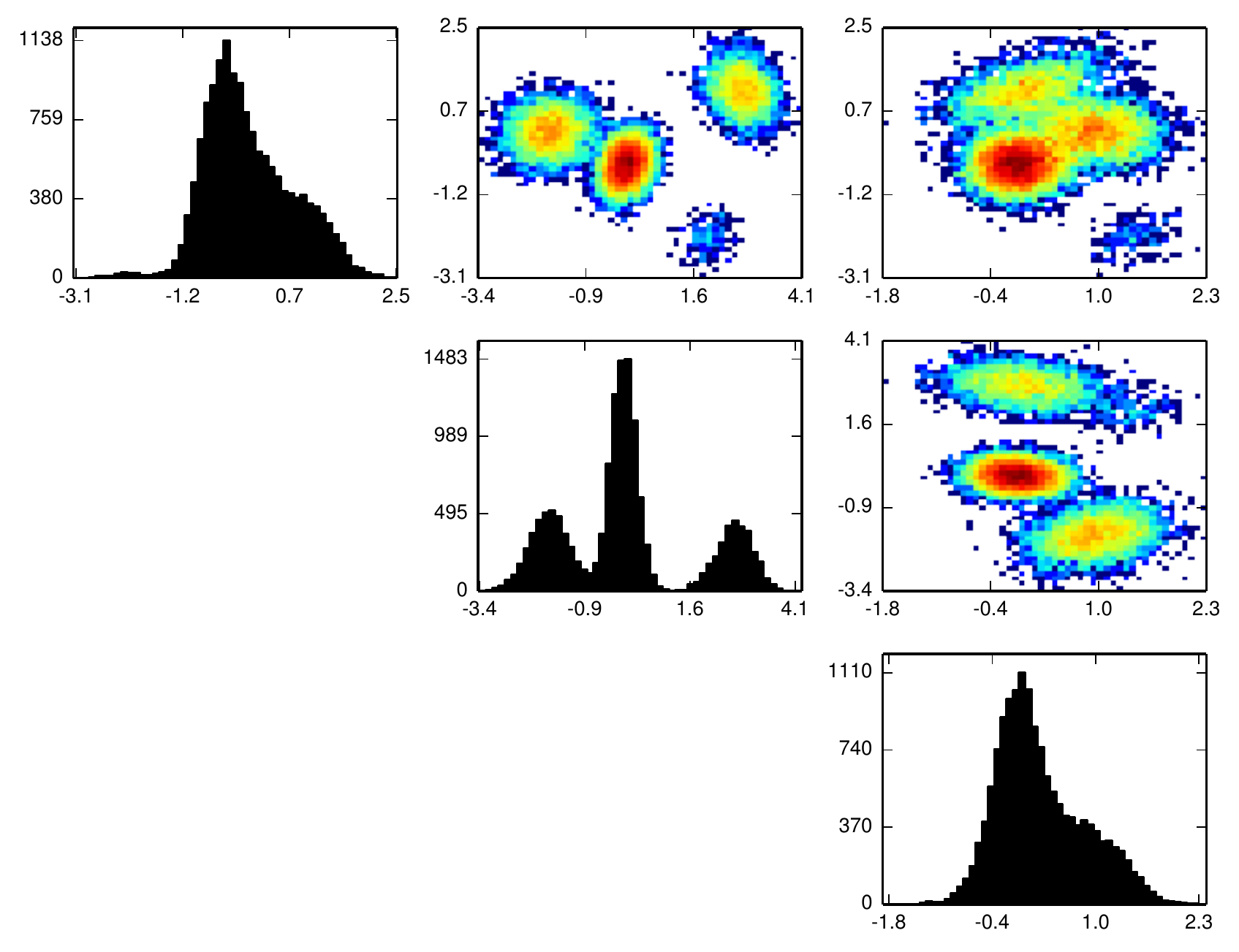}\\
			(a)
		\end{minipage}
		\begin{minipage}[b]{0.45\textwidth}
			\centering
			\includegraphics[width=\textwidth]{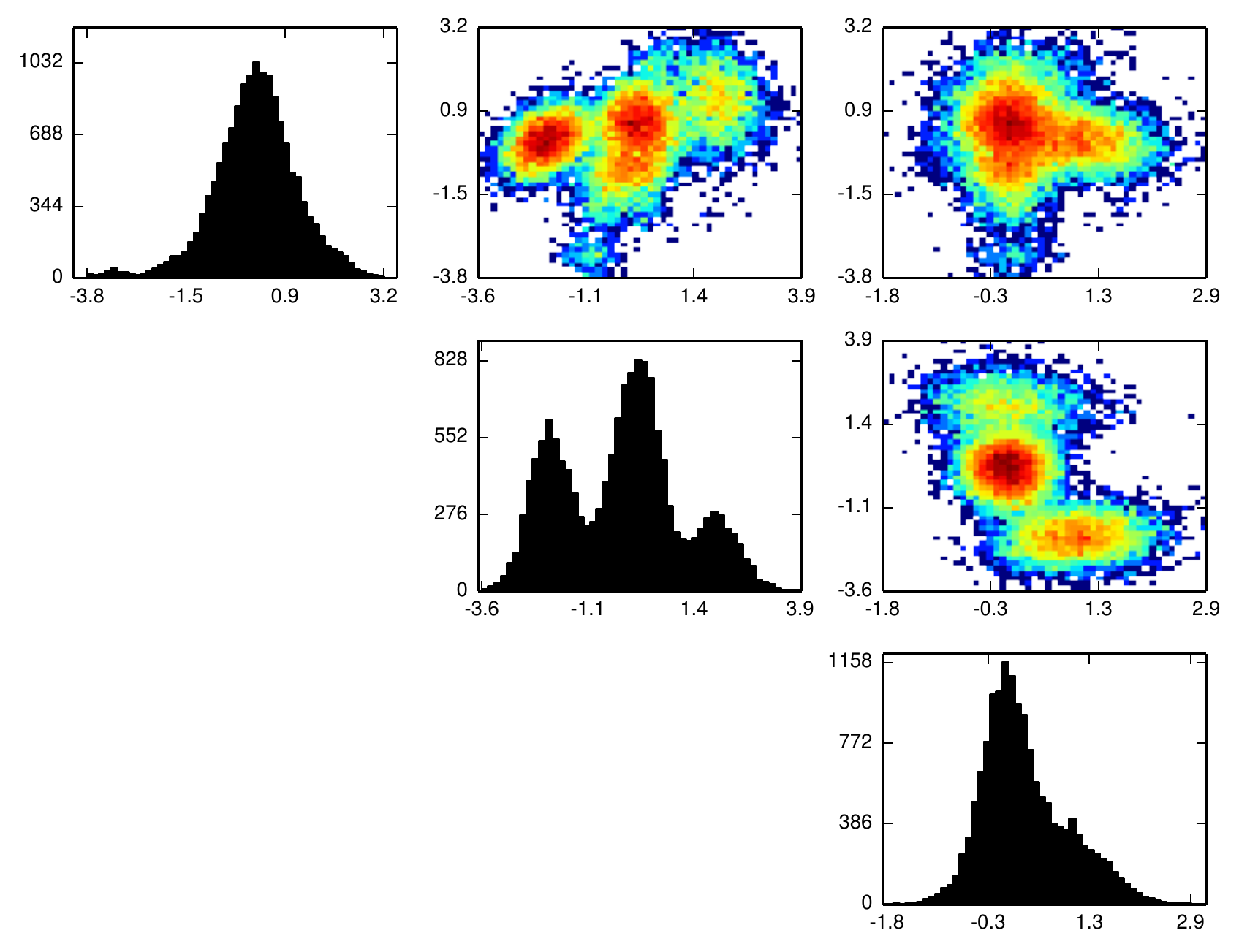} \\
			(b)
		\end{minipage}
	\end{center}
	\caption{One and two dimensional histograms of $\num{15000}$ posterior draws of $\mv{Y}$ for the flow cytometry sample displayed in Fig.\ \ref{fig:hist} (a).   In (b) we show  $\num{15000}$ posterior draws of $\mv{Y}$ drawn uniformly from all the flow cytometry samples, thus matching Fig.\ \ref{fig:hist} (b). } 
	\label{fig:hist_post}
\end{figure}

\begin{figure}[tbp]
	\centering
	\includegraphics[scale=0.5]{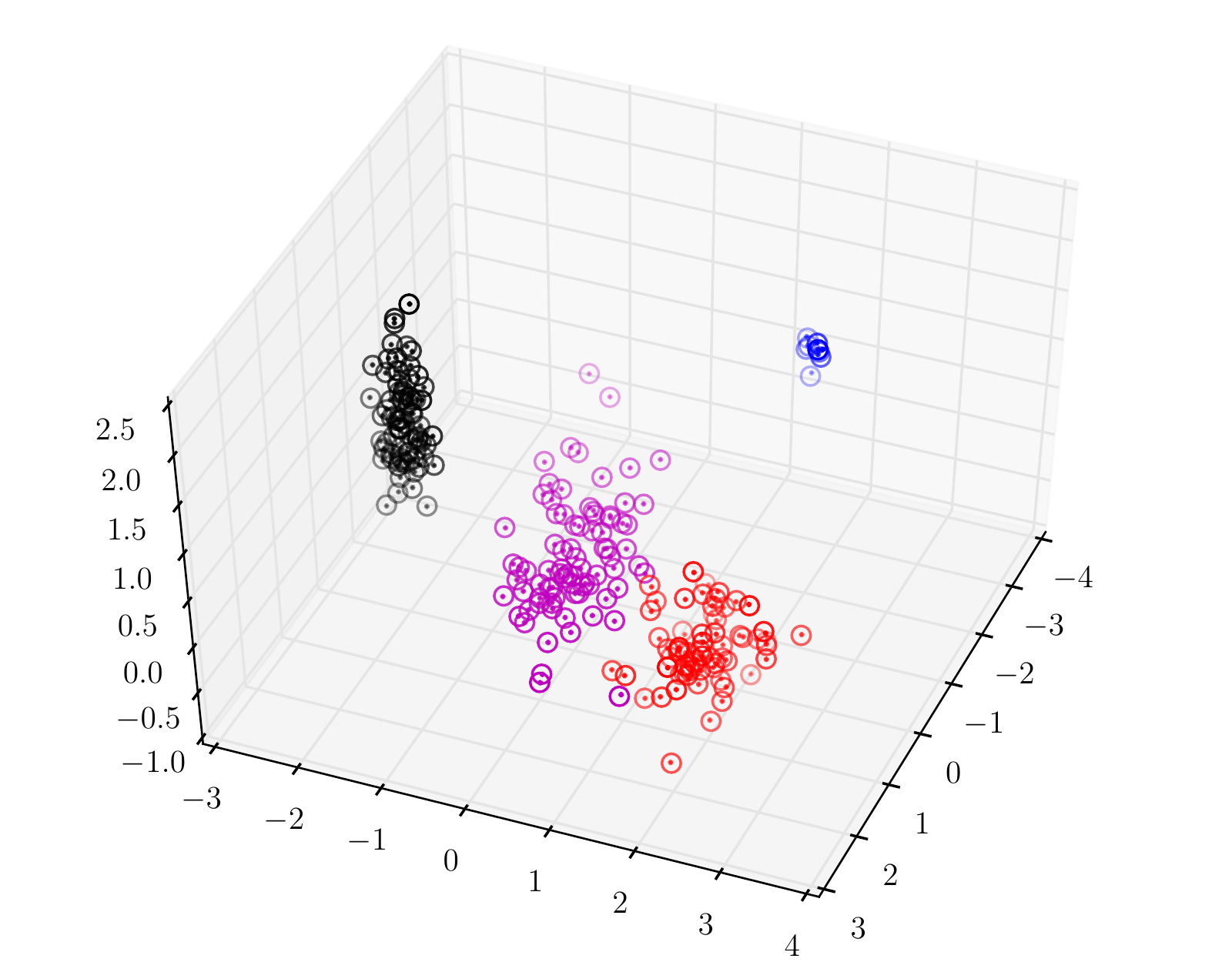}
	\caption{The posterior mean of the clusters centers, $\mv{\mu}_{jk}$ (dots), and the true cluster centers (circles).}
	\label{fig:clustcenters}
\end{figure}

The priors are chosen to be non-informative, the prior parameters as well as the initial values for the MCMC sampler are given in the Supplemental material, Appendix \ref{sec:simgen}. The posterior distribution is explored by generating samples of the parameters in $10^5$ iterations, after a burn-in period of $10^4$ iterations from the MCMC sampler. In each iteration we also generate a sample of $\mv{Y}$, i.e.\ a sample from the posterior predictive. In Fig.\ \ref{fig:hist_post} we mimic the plots in Fig.\ \ref{fig:hist}, with the samples coming from the posterior predictive distribution of $\mv{Y}$. 
Fig.\ \ref{fig:clustcenters} displays dots at the posterior mean locations of the mixture component means $\mv{\mu}_{jk}$ whose posterior probability of being active is greater than $1\%$; the true locations of the active clusters are displayed as circles. The model is able to detect which clusters that are active and which are not, and to find the location of the component means.

\begin{figure}[tbp]
	\centering
	\includegraphics[width=.6\textwidth]{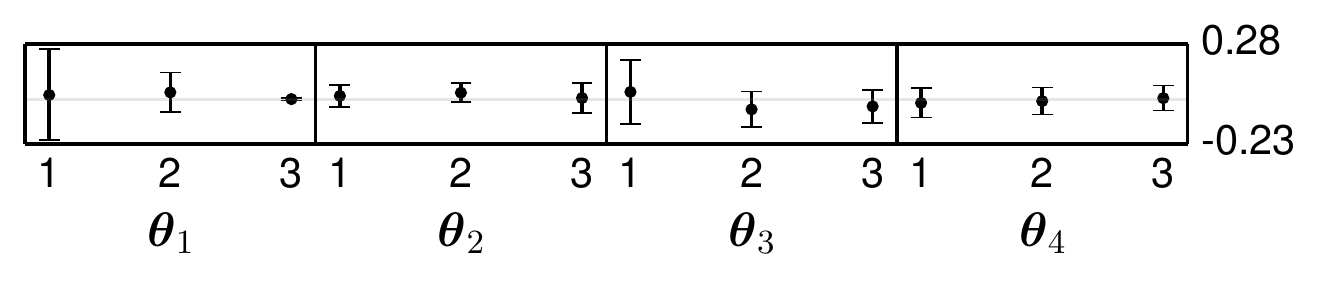}
	\caption{The difference between the true value of each entry in each $\mv{\theta}_k$ and the approximated marginal posterior distribution generated by the MCMC sampler. The black dot represents the median and the vertical line goes between the $2.5\%$ and $97.5\%$ quantiles. The light gray horizontal line is the $0$ line.} 
	\label{fig:thetasim}
\end{figure}

\begin{figure}[tbp]
	\centering
	\begin{center}
		\begin{minipage}[b]{.45\linewidth}
			\centering
			\includegraphics[width=\textwidth]{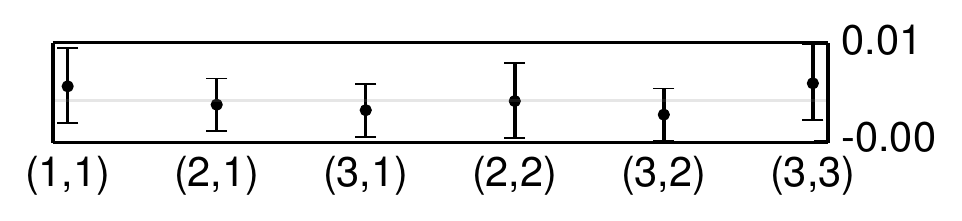}\\
			(a) $\mv{\Psi}_1/(\nu_1-4)$
		\end{minipage}
		\begin{minipage}[b]{.45\linewidth}
			\centering
			\includegraphics[width=\textwidth]{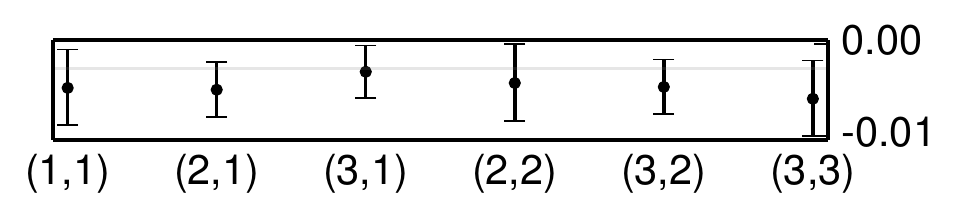} \\
			(b)$\mv{\Psi}_2/(\nu_2-4)$
		\end{minipage}
		\begin{minipage}[b]{.45\linewidth}
			\centering
			\includegraphics[width = \textwidth]{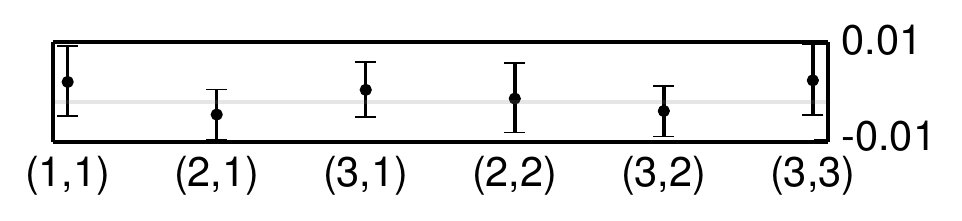} \\
			(c) $\mv{\Psi}_3/(\nu_3-4)$
		\end{minipage}
		\begin{minipage}[b]{.45\linewidth}
			\centering
			\includegraphics[width = \textwidth]{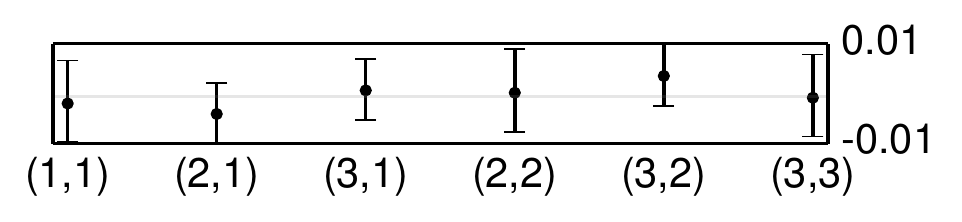} \\
			(d) $\mv{\Psi}_4/(\nu_4-4)$
		\end{minipage}
	\end{center}
	\caption{The difference between the true value of each of the entries in $\mv{\Psi}_k/(\nu_k - 4)$ and the approximated marginal posterior distribution generated by the MCMC sampler. 
		The black dot shows the median, and the black vertical line goes between the $2.5\%$ and $97.5\%$ quantiles. The light gray horizontal line is the $0$ line.} 
	\label{fig:Sigmasim}
\end{figure}

Finally in Fig.\ \ref{fig:thetasim} and \ref{fig:Sigmasim}, the marginal posterior distributions of the latent cluster parameters $\mv{\theta}_k$ and $\mv{\Psi}_k$, subtracted by their true values, are presented. In Fig.\ \ref{fig:thetasim} the dot represents the difference between the median of posterior distribution and the true value of each $\mv{\theta}_k$. The vertical lines represent the $2.5\%$ and $97.5\%$ quantiles. Fig.\  \ref{fig:Sigmasim} displays results for each latent covariance matrix $\mv{\Psi}_k/(\nu_k - 4)$ in an analogous way. From Fig.\ \ref{fig:thetasim} and Fig. \ref{fig:Sigmasim} we see that the true parameters of both the means and the covariances are all between the 2.5\% and 97.5\% quantiles of the posterior distribution. 

% % % % %
%
% Real data
%
% % % % % % %

\subsection{Flow cytometry data} \label{sec:realworld}
We use our model to study lymphocyte cell populations among peripheral mononuclear blood cells (PMBC) from four healthy donors. The data has previously been used in the evaluation of a population matching algorithm (\citealp{azad13}) and is available from the R package \texttt{healthyFlowData}. Lymphocytes are the leading actors in the adaptive immune response and also play a role in the innate immune response. It is well known that the sizes of the lymphocyte subpopulations vary between individuals (\citealp{jentsch05}).

\begin{figure}[tbp]
\centering
	\begin{minipage}[b]{.48\textwidth}
		\centering
		\includegraphics[width = \textwidth,trim = 0mm 0mm 0mm 0mm, clip]{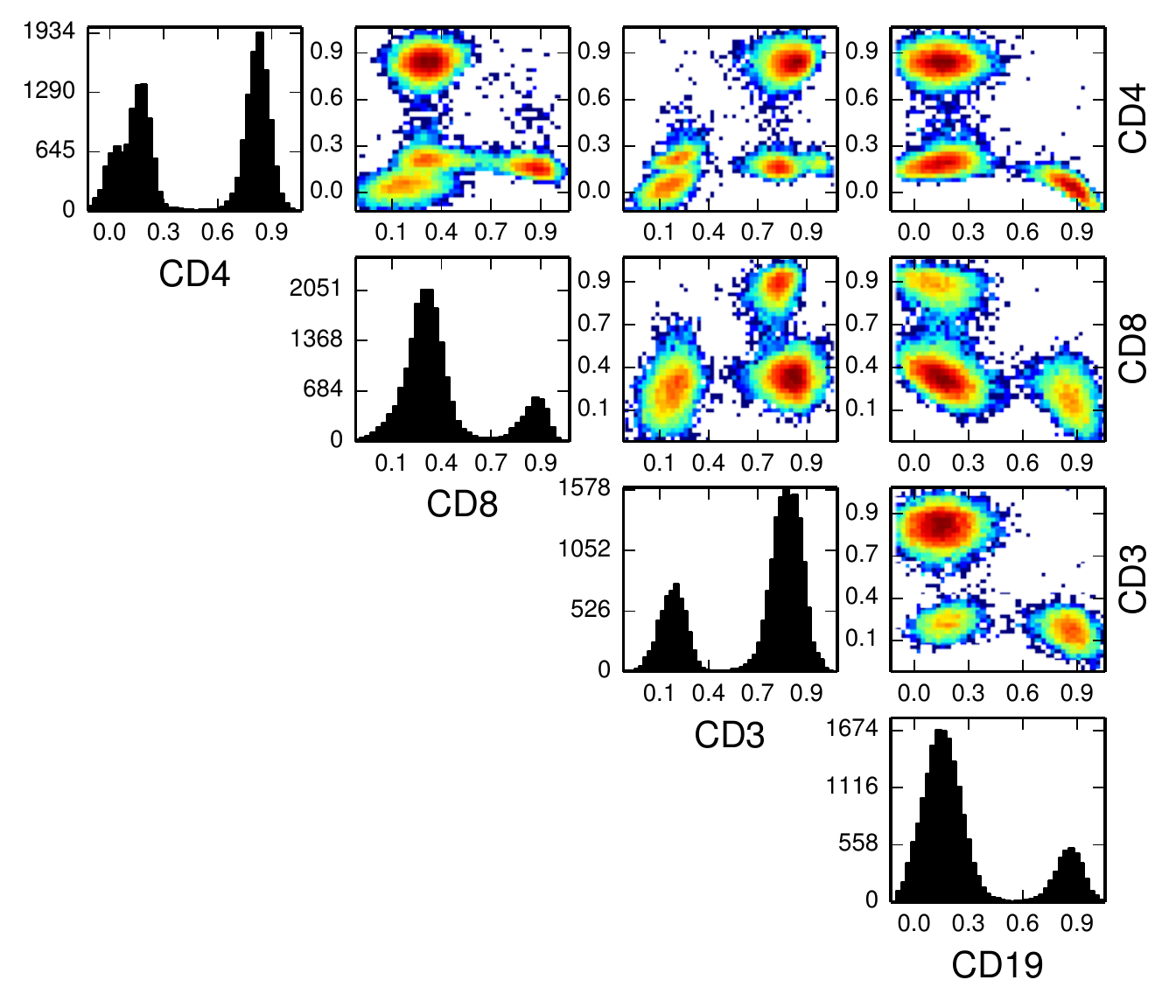}
		(a)
	\end{minipage}
	\begin{minipage}[b]{.48\textwidth}
		\vspace{10pt}
		\centering
		\includegraphics[width = \textwidth,trim = 0mm 0mm 0mm 0mm, clip]{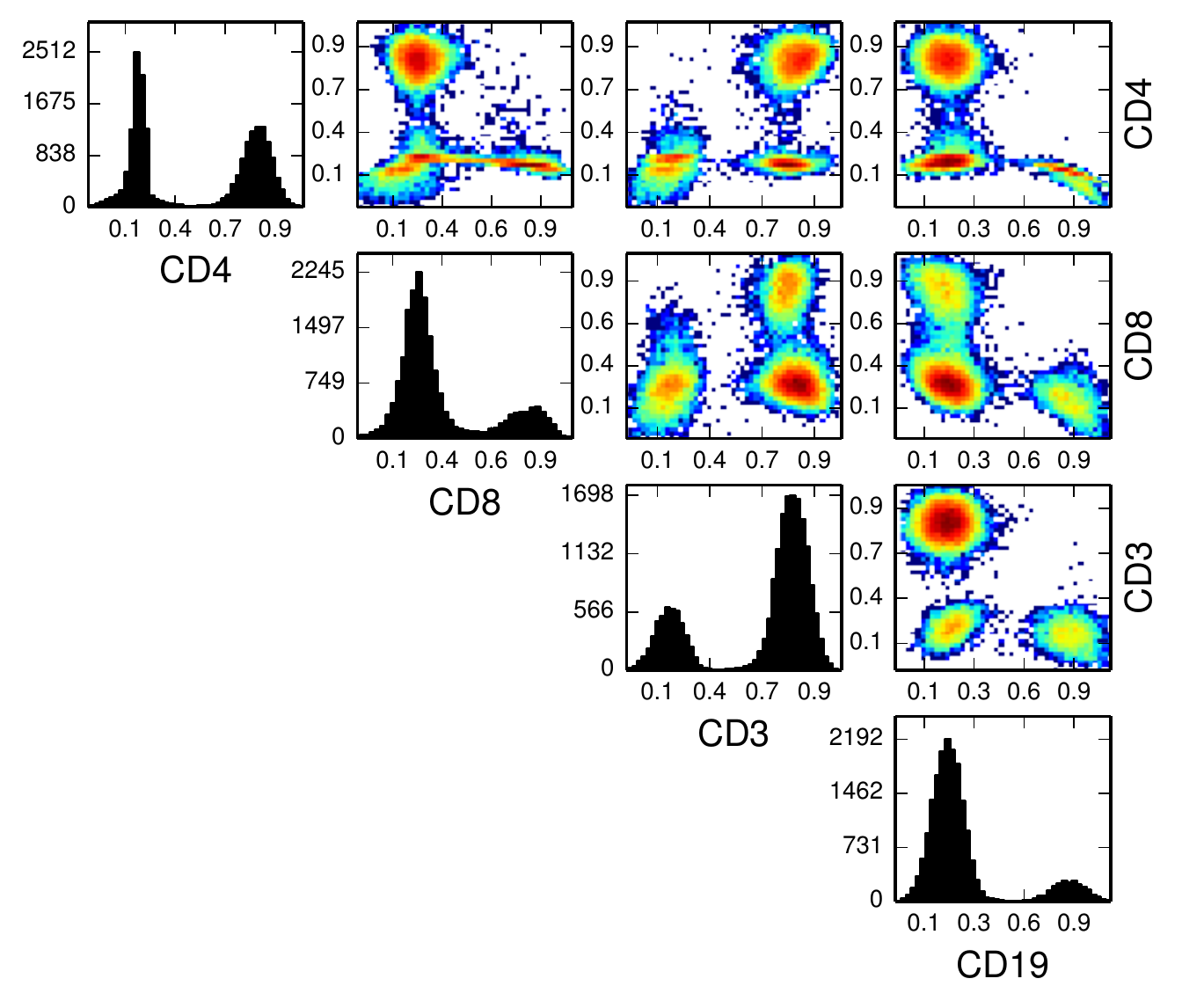}
		(b)
	\end{minipage}
	\caption{One and two dimensional histograms of flow cytometry data. (a) One of the samples from healthyFlowData, with \num{19160} data points. (b) \num{19160} data points drawn from the twenty samples in healthyFlowData pooled together.}
	\label{fig:truehistrw}
\end{figure}

The blood from each of the four donors was split into five parts and then analyzed, the data set has therefore twenty flow cytometry samples. The data in \texttt{healthyFlowData} has been preprocessed as detailed in the Supplemental material, Appendix \ref{sec:datadetails}. As an additional preprocessing step we scaled it using the 1\% and 99\% percentiles $q_{0.01}$ and $q_{0.99}$ of the pooled data, with the same scaling for all samples, so that $q_{0.01} = 0$ and $q_{0.99} = 1$ for each marker for the pooled data. Univariate and bivariate histograms of one of the samples as well as all samples pooled together are shown in Fig.\ \ref{fig:truehistrw}.

The data set has measurements of the four markers CD4, CD8, CD3 and CD19. If a cell population has a high expression of a certain marker, say CD4, it is said that it is positive for this marker, denoted CD4+. A population with low expression is called negative and if the expression is intermediate it is called dim or bright. It is well known that using the four markers above, five types of lymphocytes can be distinguished (\citealp{murphy12}). CD19 is a pan B-cell marker, i.e.\ it is highly expressed on all B-cells, and CD3 is a pan T-cell marker. The lymphocytes with low expression of CD3 or CD19 are natural killer (NK) cells. There are two major cell populations among the T cells, namely helper T cells and cytotoxic T cells. The helper T cells are CD4 positive and CD8 negative whereas the cytotoxic T cells are CD8 positive and CD4 negative. There is also a CD4--CD8-- population comprising mainly so called $\gamma\delta$ T cells (\citealp{gertner07}). Table \ref{tab:lymph} summarizes these subpopulations; we want to follow them in the twenty samples, while at the same time accounting for unknown populations due to incomplete biological knowledge or technical artifacts.

\begin{table}[tbp]
	\centering
		\caption{Lymphocyte subpopulations.}
			\label{tab:lymph}
	\begin{tabular}{lcccc}
		\toprule
		& CD4 & CD8 & CD3 & CD19 \\ \midrule
		B cells 	& 	&	& -- & + \\
		Helper T cells	& + & -- & + & -- \\
		Cytotoxic T cells & -- & + & + & -- \\
		CD4--CD8-- T cells & -- & -- & + & -- \\
		NK cells & & & -- & -- \\ \bottomrule
	\end{tabular}
\end{table}

Often in Bayesian analysis of mixture models one tries to construct as non-informative priors as possible (\citealp{RJMIX1997},\citealp{roeder97}). However, in our setup we have two kinds of prior information that we want to utilize. First, we have semi-quantitative information about marker expression in the previously known populations described in Table \ref{tab:lymph}. Even though there is no reference value for marker expression of a positive population (\citealp{shapiro05})---since we have positive and negative populations present for each of the four markers in the data---we can set relative values for the prior parameters $\mv{t}_k$ based on the scaling of the data. Second, since the sample components linked to one latent cluster are supposed to represent one and the same cell population, the spread of their means must be reasonable. That is, for any latent cluster, the sample component means linked to it should be closer to its own mean than to the means of latent clusters that do not represent the same population, i.e.\ that are not merged with it during post-processing. This second piece of information is used to set $n_\theta$, $n_\Psi$ and $\mv{Q}_k$. It is harder to translate to precise values of the parameters, but it can easily be detected if the priors allow too much variation. The chosen parameters as well as the initialization procedure are described in the Supplementary material Appendix \ref{sec:priors}.

Apart from the five latent clusters with informative priors on their latent means we also have a number of additional latent components with non-informative priors on their means. These additional components can capture outliers and unknown populations, but they can also capture parts of the known populations, especially if the populations are non-Gaussian and hence cannot be well approximated by one Gaussian component. We merge components with sufficient overlap as described in Section \ref{sec:merging} after inferring the model to obtain the populations in one piece. With a higher number of components the data set can be described more accurately, but the interpretation of each component gets less clear. We use in total $K = 17$ components since we found that this was sufficient to describe the data set well, in the sense that one- and two-dimensional marginal distributions are accurate.

With these priors we get in most cases appropriate variation between components when rerunning the analysis below, however sometimes (two out of six cases) the variation in the location of one of the seventeen clusters is a bit too high.

We run $10^5$ burn-in iterations of our algorithm and then $R = 10^5$ production iterations to get samples of
$\mv{\Theta}^{(r)} = \left(\mv{\mu}_{jk}^{(r)}, \mv{\Sigma}_{jk}^{(r)}, \mv{\theta}_k^{(r)}, \mv{\Psi}_k^{(r)},  \nu_k^{(r)}, \mv{\pi}_j^{(r)}\right), \, r = 1,\dotsc,R
$
from the posterior distribution. Convergence of the MCMC sampler is established using trace plots, displayed in the Supplementary material, Appendix \ref{sec:evaluation}. We use the means of $\mv{\mu}_{jk}^{(r)}, \mv{\Sigma}_{jk}^{(r)}, \mv{\theta}_k^{(r)}$ and $\mv{\Psi}_k^{(r)}/(\nu_k^{(r)}-d-1)$ to get point estimates of sample component and latent cluster means and covariance matrices; the means of $\mv{\pi}_j^{(r)}$ are used to get point estimates of the mixing proportions. 
Furthermore, in each iteration we draw one synthetic cell measurement from the models of selected flow cytometry samples and one from the model of the pooled data, in order to evaluate how well the model fits the data. 

\begin{figure}[tbp]
\centering
	\begin{minipage}[b]{.45\linewidth}
		\centering
		\includegraphics[width = \textwidth,trim = 0mm 0mm 0mm 0mm, clip]{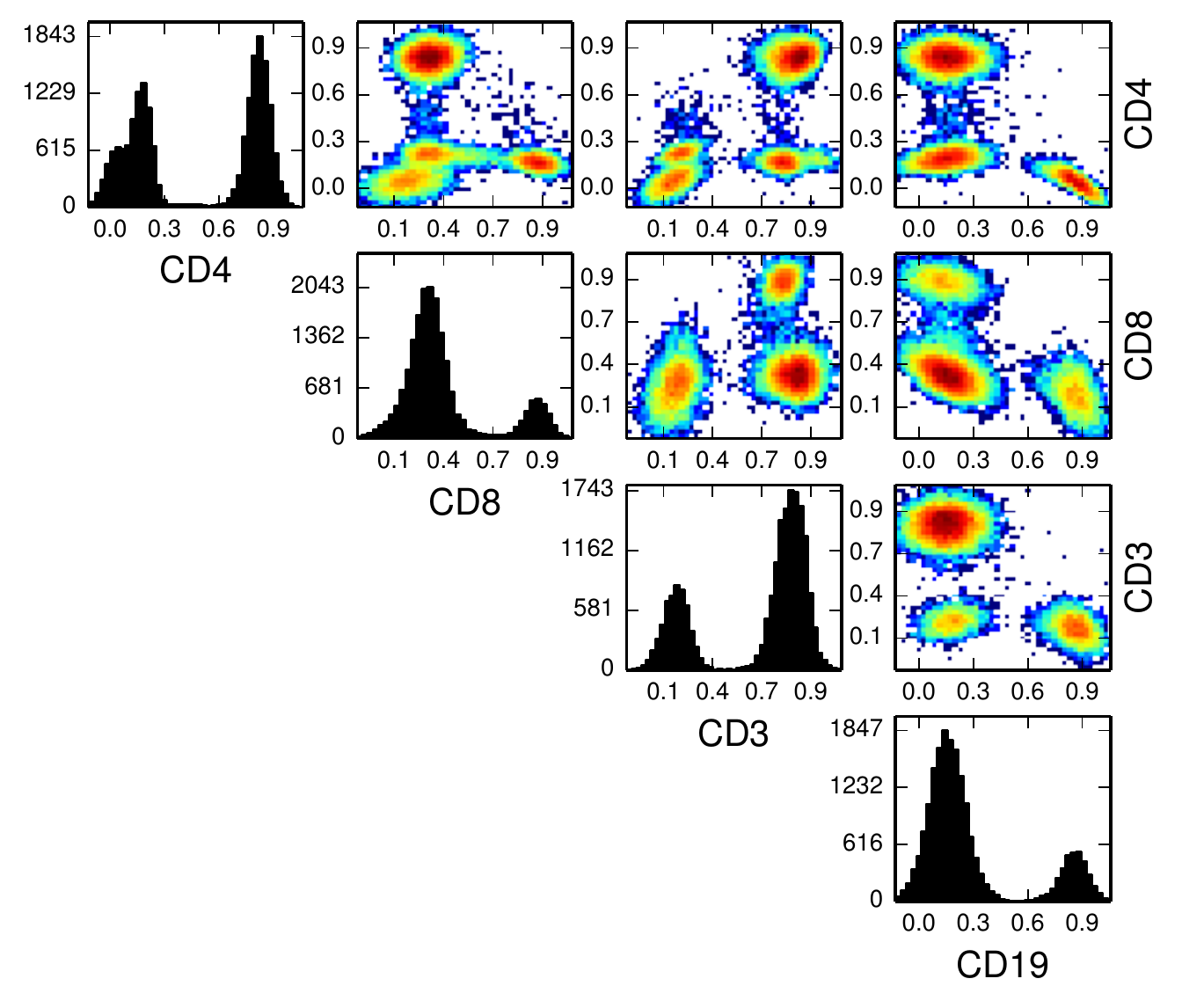}
		(a)
	\end{minipage}
	\begin{minipage}[b]{.45\linewidth}
		\centering
		\includegraphics[width = .9\textwidth]{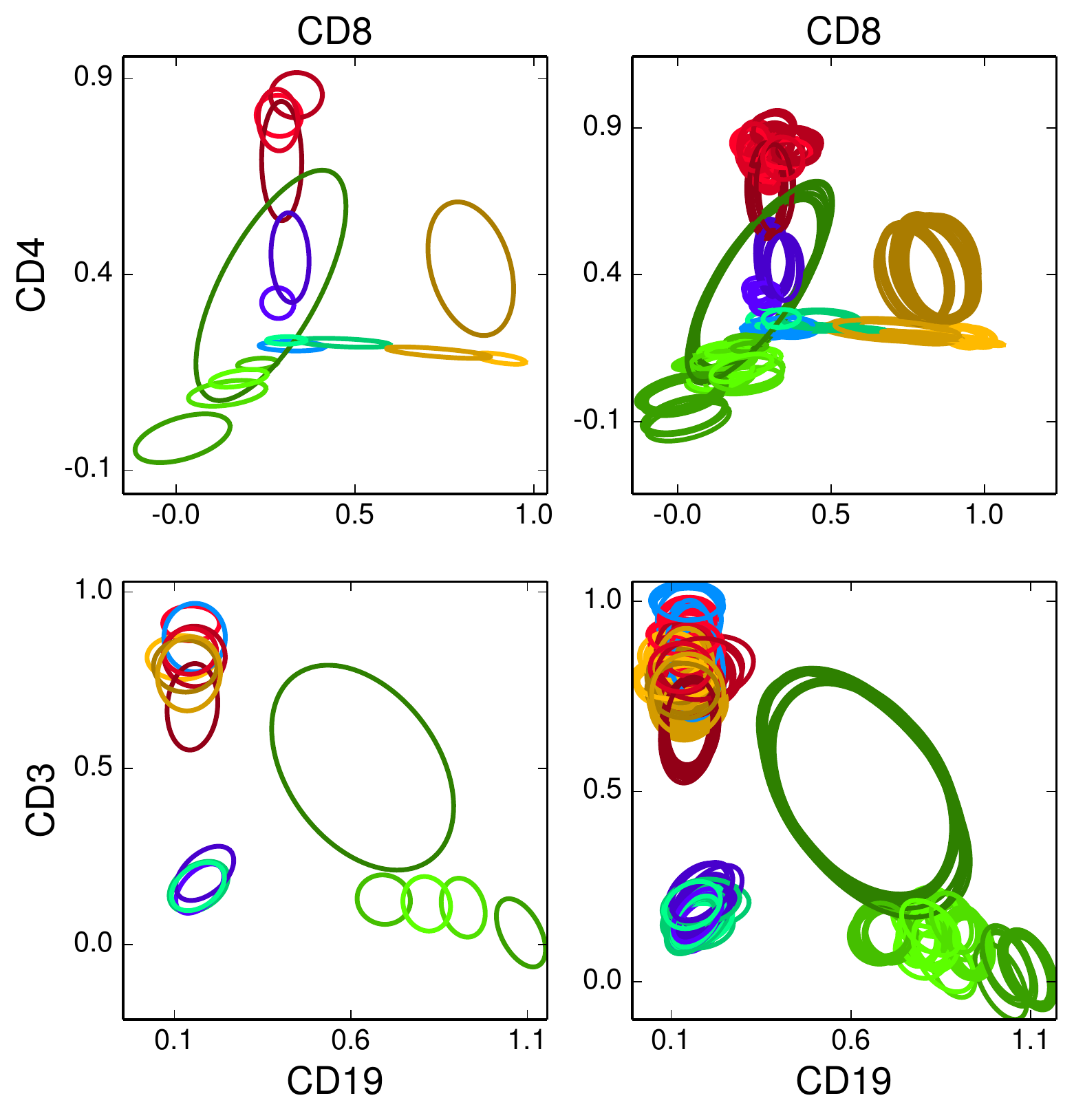}
		(b)
	\end{minipage}
	\caption{Bayesian hierarchical model of lymphocyte subsets of PBMC from healthy individuals. (a) One and two dimensional histograms of \num{19160} synthetic data points generated from the inferred model of the flow cytometry sample depicted in Fig.\ \ref{fig:truehistrw} (a). (b) Component parameter representations of inferred latent clusters and mixture components across the twenty flow cytometry samples. The center of each ellipse is the mean and each semi-axis is an eigenvector with length given by the corresponding eigenvalue of the projected covariance matrix. In the left column parameters of latent clusters $(\mv{\theta}_k,\frac{1}{(\nu_k-d-1)}\mv{\Psi}_k)$ are shown, in the right column parameters for the mixture components in each sample  $(\mv{\mu}_{jk},\mv{\Sigma}_{jk})$ are shown. Each component or cluster is depicted with the same color as in Fig.\ \ref{fig:HFdiagn}; different shades of same color corresponds to latent clusters that have been merged. }
	\label{fig:HFresInf}
\end{figure}

Fig.\ \ref{fig:HFresInf} (a) shows univariate and bivariate histograms of the synthetic cell measurements for one flow cytometry sample. Fig.\ \ref{fig:truehistrw3} and Fig.\ \ref{fig:truehistrwpooled} in the  Supplementary material display histograms of synthetic measurements of another sample as well as of the pooled data. From these results it is clear that the inferred model is accurate and captures the variation across samples, which a model of pooled data cannot do.

\begin{figure}[tbp]
\centering
	\begin{minipage}[t]{.28\textwidth}
		\centering
		\vspace{0pt}
		\includegraphics[scale=.6,trim = 0mm 0mm 0mm 0mm, clip]{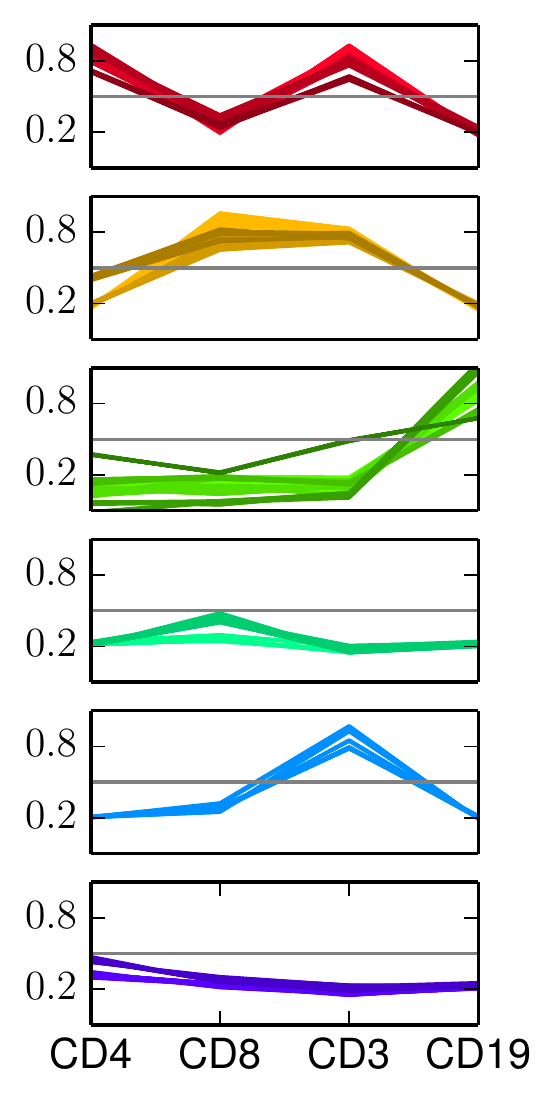}
		(a)
	\end{minipage}
	\begin{minipage}[t]{.27\textwidth}
		\centering
		\vspace{0pt}
		\includegraphics[scale=.6,trim = 0mm 0mm 0mm 0mm, clip]{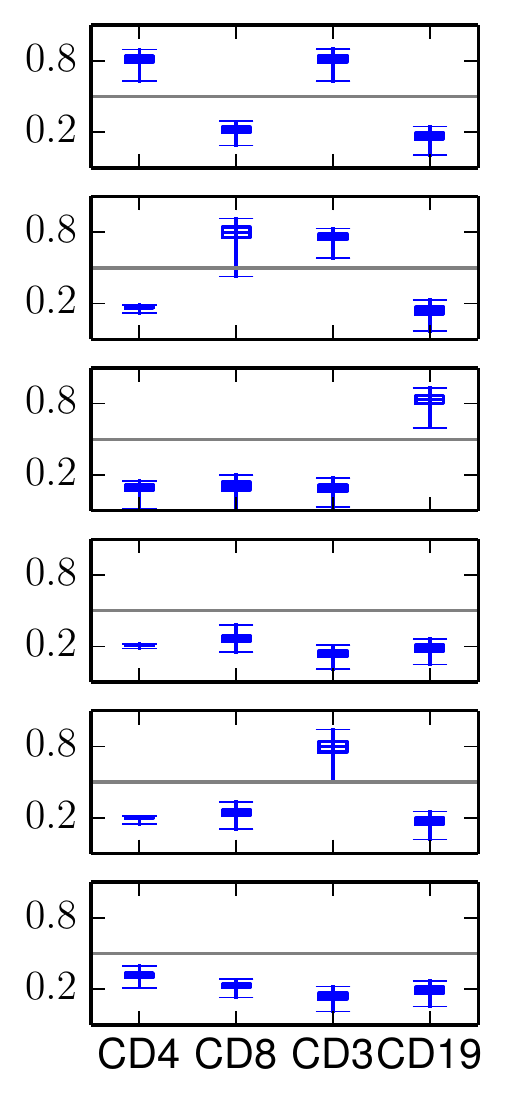}
		(b)
	\end{minipage}
	\begin{minipage}[t]{.27\textwidth}
		\centering
		\vspace{0pt}
		\includegraphics[scale=.6,trim = 0mm 0mm 0mm 0mm, clip]{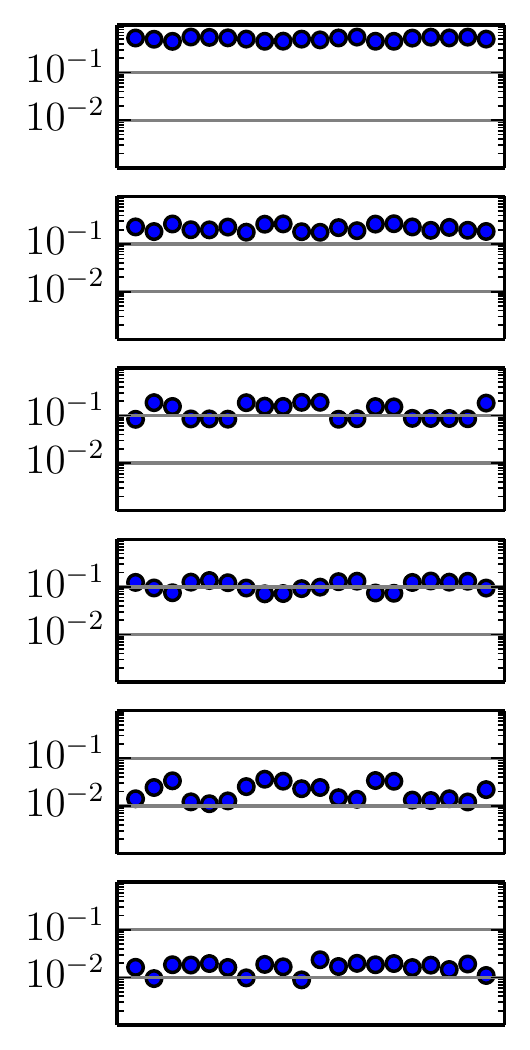}
		\vspace{0mm}
		
		(c)
	\end{minipage}
	\caption{Summary statistics of the six inferred cell populations, ordered by population size. (a) Component mean locations for each population. Different shades of a color in the same plot corresponds to different latent components form the same super component. Populations: CD4 T-cells (red), CD8 T-cells (yellow), B-cells (green), monocytes or NK cells (turquoise), CD4-CD8-- T-cells (blue), and monocytes or NK cells (purple). (b) Boxplots of the soft clusters. The boxes go between $q_{km,0.25}$ and $q_{km,0.75}$ and the whiskers extend to $q_{km,0.01}$ and $q_{km,0.99}$. The $\alpha$-quantile for (merged) component $k$ in dimension $m$, $q_{km,\alpha}$, is here defined as $q_{km,\alpha} = \min_{i'j'}\{Y_{i'j'm} : \alpha < \sum_{ij:Y_{ijm} < Y_{i'j'm}} w_{ijk} \}$.
		(c) Population proportions for each of the flow cytometry samples.}
	\label{fig:HFdiagn}
\end{figure}

Merging of latent components results in six cell populations, summary statistics of these are shown in Fig.\ \ref{fig:HFdiagn}. Two of the populations fit the NK cell expression pattern in Table \ref{tab:lymph}; there are multiple possible explanations of this. One explanation is be that one of the population is monocytes, which have front and side scatter properties similar to lymphocytes (\citealp{BD00})---making it plausible that some of them have been included---and also are CD3--CD19--. Another explanation could be that there are two NK populations which are differentially activated. There are strong evidence that there are two separate populations, see for instance the histograms of their joint CD4 expression shown in Fig.\ \ref{fig:NKhist} in the Supplementary material.

The component representing outliers has only a very small proportion of the cells assigned to it, the median of $\hat \pi_{j0}$ across samples is 0.0003, the maximum is 0.0011.

\begin{figure}[tbp]
\centering
	\includegraphics[width = .6\textwidth,trim = 0mm 0mm 0mm 0mm, clip]{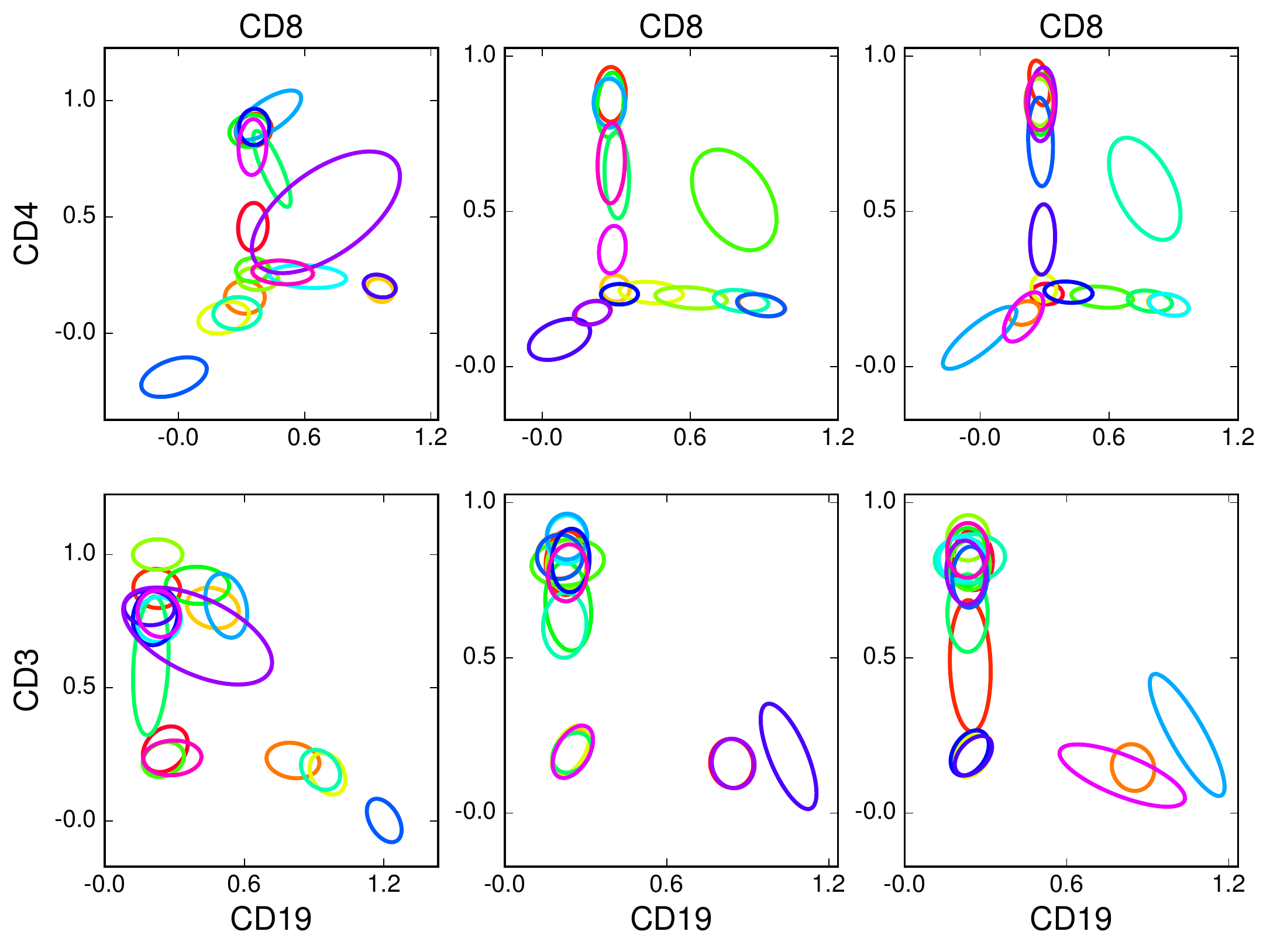}
	\caption{Component parameter representations of inferred mixture components in independent Gaussian mixture models of three flow cytometry samples. The two samples depicted in the two right columns are technical replicates. Note that there is no correspondence between colors between columns.}
	\label{fig:EM}
\end{figure}

The variation between flow cytometry samples is systematized in the hierarchical model, results of this can be seen in Fig.\ \ref{fig:HFdiagn} (a) and Fig.\ \ref{fig:HFresInf}. For comparison, we fit Gaussian mixture models using the expectation-maximization algorithm to each flow cytometry sample separately. In this case there are no clear correspondences among the mixture components between samples, as seen in Fig.\ \ref{fig:EM}. When the data set was studied previously with an algorithm matching populations found by separate analysis of the samples, this was only done with a coarse partition of the cell measurements, with four cell populations (\citealp{azad13}).

In downstream analysis of flow cytometry data the locations and shapes of the cell populations as well as their sizes can be used. For example, one cell population might have varying expression of one marker in different samples; this can be studied using the mean parameters $\mv{\mu}_{jk}$. We use cell population sizes to show how technical variation between replicates can be separated from biological variation between donors.

\begin{figure}[tbp]
	\centering
	\includegraphics[width = .5\textwidth,trim = 0mm 0mm 0mm 0mm, clip]{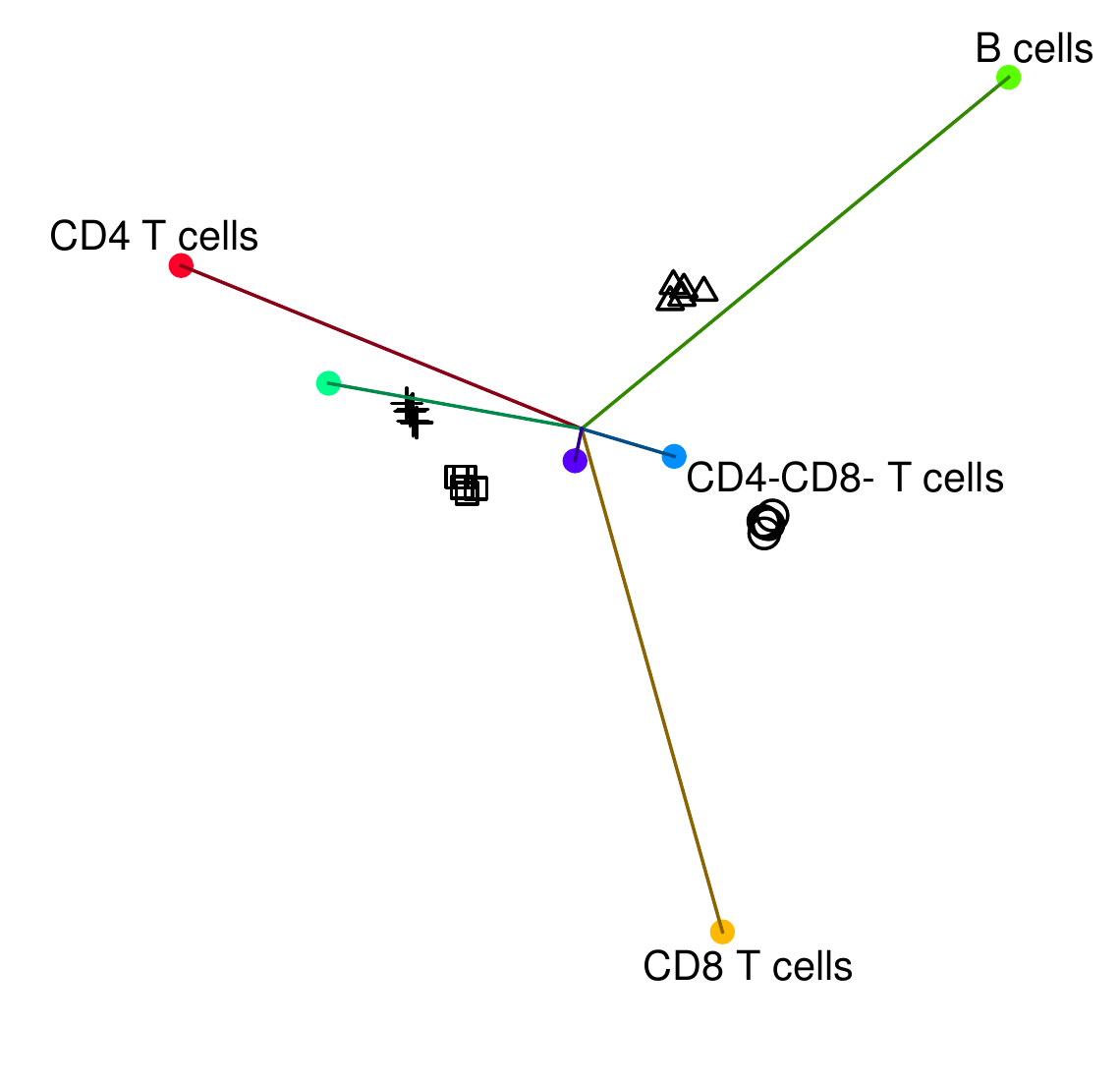}
	\caption{PCA biplot. Colors for cell populations are the same as in Fig.\ \ref{fig:HFdiagn}. Samples from the same donor are plotted with the same marker. Samples from different donors are well separated.}
	\label{fig:pca}
\end{figure}

We estimate the size of each cell population in each sample by sum of the mixing proportions for the sample components representing that population. We then do a principal component analysis on the population sizes. Almost all of the variance, 99.3\%, is captured in the first two principal components. A biplot onto the two first principal components is shown in Fig.\ \ref{fig:pca}. We see that the biological variation between donors is much larger than the technical variation between replicates, samples from different donors are well separated.

% % % % % % % % % % % % % % % % %
%   Discussion
% % % % % % % % % % % % % % % % % 
\section{Discussion} \label{sec:disc}
In this paper we have presented a new Bayesian hierarchical model designed for joint population identification in many flow cytometry samples. The model captures the variability in shapes and locations of the populations between the samples and has the possibility to include expert knowledge. We showed that for a synthetic data set generated from the model, the parameters were recovered with high accuracy through a MCMC sampling scheme. The sampling scheme was then applied to a real flow cytometry data set which contained five populations whose expression patterns were well-known to experts. After merging latent clusters using criteria based on Bhattacharyya distance and Hartigan's dip test, six populations were obtained---apart from the previously known populations, one additional population which could either be monocytes or NK cells.

How much clusters should be merged is however a decision that needs to be taken by the interpreter of the data. The criteria we have used should be taken as guidelines. Other merging criteria, for example directly estimated misclassification probabilities (\citealp{hennig10}), and diagnostic plots which evaluate the separation of clusters could also be utilized to guide such decisions. As an example, in the analysis of the data set studied in this article, a flow cytometry data analyst might want to consider the small cluster which is CD8+ and has intermediate CD4 expression to be separate from the CD8 T-cells.

For very large flow cytometry data sets the running time can be prohibitive and downsampling is then required. A direction for future research is to find ways to downsample so that also small cell populations can be resolved, as has been done for the analysis of single flow cytometry samples (\citealp{naim14}).

The priors that we used in the real data experiments were rudimentary; incorporating more detailed knowledge about cell populations could speed up convergence. With smart ways of formulating priors based on populations found in other flow cytometry data sets---possibly with different sets of markers than the data set under study---our method could lead the way to incremental learning of flow cytometry populations where well-known populations will be easily characterized.

There are interesting approaches to analysis of multiple flow cytometry data samples in parallel which are not based on clustering (\citealp{aghaeepour13}). But the power of clustering approaches is to provide a comprehensive description of all cell populations in the samples which can reveal subtle signals and systematic perturbations. Finding good clustering algorithms which can serve as reliable automatic gating methods is therefore an important task. We feel strongly that analyzing many samples jointly, and using expert prior knowledge when doing automated gating is an important requirement to get the high performance which is needed.

% % % % % % % % % % % % % % % % %
%   Software
% % % % % % % % % % % % % % % % % 
\section{Software}
Our implementation of the MCMC sampling is available as a Python package at \url{https://github.com/JonasWallin/BayesFlow}. Its parallel implementation is based on OpenMPI through the Python package mpi4py.

\section{Supplementary material}
Supplementary material is available. It contains the posterior, the MCMC sampling scheme, additional details on the merging of components, information about the data generation, priors and initialization for the synthetic data example; additional details on the flow cytometry data set, the priors and the initialization procedure used when studying this data set and further results pertaining to the flow cytometry data set.

% % % % % % % % % % % % % % % % %
%   Acknowledgements
% % % % % % % % % % % % % % % % % 
\section{Acknowledgements}
We would like to thank Ariful Azad for sharing the flow cytometry data set and for providing information about the preprocessing of the data. We would also like to thank Alejandra Urrutia at Institut Pasteur, Paris, for assisting with the interpretation of the data.

The first author is supported by Knut and Alice Wallenbergs stiftelse.

\bibliography{bayesianmodel,FlowCyto}

\end{document}